\documentclass[aps,pra,reprint,superscriptaddress,footinbib,]{revtex4-2}

\usepackage[T1]{fontenc}			
\usepackage[utf8]{inputenc}			
\usepackage[english]{babel}			
\usepackage[protrusion=false]{microtype}	
\usepackage{comment}
\usepackage{xcolor} 						
\definecolor{red}{rgb}{1,0,0}				
\definecolor{blue}{rgb}{0,0,1}				
\definecolor{black}{rgb}{0,0,0}				

\usepackage{soul} 
\definecolor{hlyellow}{rgb}{0.95,0.95,0}
\definecolor{hlgreen}{rgb}{0,0.95,0}
\sethlcolor{hlyellow}
\soulregister \ref 7
\soulregister \cite 7
\soulregister \citet 7
\soulregister \footnote 8
\soulregister {\ } 7
\soulregister \figref 7
\soulregister \tabref 7
\soulregister \secref 7
\soulregister \supref 7
\soulregister \refref 7
\soulregister \eqref 7
\soulregister \supmat 7

\usepackage{siunitx}
\usepackage{amsmath} 
\usepackage{amssymb,amsfonts,mathrsfs} 
\usepackage{array, dcolumn} 

\usepackage{graphicx} 

\usepackage{hyperref} 
\usepackage{cleveref}

\definecolor{dullmagenta}{rgb}{0.4,0,0.4} 
\definecolor{darkblue}{rgb}{0,0,0.4}
\definecolor{medblue}{rgb}{0,0,0.6}
\definecolor{lightblue}{rgb}{0,0,0.8}
\hypersetup{
	colorlinks=true, 		
	breaklinks=true,		
	linkcolor=lightblue,
	citecolor=lightblue,
	urlcolor=lightblue,
	filecolor=dullmagenta
}
\usepackage[all]{hypcap} 
\usepackage{physics}

\usepackage[version=4]{mhchem} 	

\newcommand{\figref}[1]{Fig.~\ref{#1}} 
\newcommand{\tabref}[1]{Tab.~\ref{#1}} 
\newcommand{\secref}[1]{Sec.~\ref{#1}} 
\newcommand{\refref}[1]{Ref.~\cite{#1}} 

\newcommand{\vPone}{V_{\overline{\mathrm{P{1}}}}}
\newcommand{\vPtwo}{V_{\overline{\mathrm{P{2}}}}}

\newcommand{\tesla}{\ \text{T}}      				

\newcommand{\diff}{\mathrm{d}}

\newcommand{\supmat}{Supplementary Material}
\newcommand{\supref}[1]{\supmat{} \secref{#1}} 

\begin{document}

\newcommand{\mytitle}
{Disentangling orbital and confinement contributions to $g$-factor in Ge/SiGe hole quantum dots}
\title{\mytitle}

\newcommand{\zrl}{IBM Research Europe -- Z{\"u}rich, S{\"a}umerstrasse 4, 8803 R{\"u}schlikon, Switzerland}
\newcommand{\yorktown}{IBM Quantum, T.\,J.\,Watson Research Center, Yorktown Heights, NY, USA}
\newcommand{\ethzurich}{Solid State Physics Laboratory, ETH Z{\"u}rich, 8093 Z{\"u}rich, Switzerland}

\author{L. \surname{Sommer}}
\email[Correspondence to: ]{lisa.sommer@ibm.com}

\affiliation{\zrl}
\author{I. \surname{Seidler}}
\affiliation{\zrl}
\author{F. J. \surname{Schupp}}
\affiliation{\zrl}
\author{S. \surname{Paredes}}
\affiliation{\zrl}
\author{N. W. \surname{Hendrickx}}
\affiliation{\zrl}
\author{L. \surname{Massai}}
\affiliation{\zrl}
\author{K. \surname{Tsoukalas}}
\affiliation{\zrl}
\author{A. \surname{Orekhov}}
\affiliation{\zrl}
\author{E. G. \surname{Kelly}}
\affiliation{\zrl}
\author{S. W. \surname{Bedell}}
\affiliation{\yorktown}
\author{G. \surname{Salis}}
\affiliation{\zrl}
\author{M. \surname{Mergenthaler}}
\affiliation{\zrl}
\author{P. \surname{Harvey-Collard}}
\affiliation{\zrl}
\author{A. \surname{Fuhrer}}
\affiliation{\zrl}
\author{T. \surname{Ihn}}
\affiliation{\ethzurich}

\date{February 27, 2026} 

\begin{abstract}
Spin qubits are typically operated in the lowest orbital of a quantum dot to minimize interference from nearby states. In valence-band hole systems, strong spin–orbit coupling links spin and orbital degrees of freedom, strongly influencing the hole $g$-factor, a key parameter for qubit control. We investigate the out-of-plane $g$-factor in Ge quantum dots using excitation (single-particle) and addition (many-body) spectra. Excitation spectra allow us to distinguish the pure Zeeman $g$-factor from orbital contributions to the magnetic field splitting of states despite the strong spin-orbit coupling. This distinction clarifies discrepancies between $g$-factors extracted with the two methods, for different orbital states and different hole numbers. Furthermore, we find gate-tunability of $g$-factors at the level of 15\%, highlighting its relevance for all-electric qubit manipulation. 

\end{abstract}

\maketitle



Spin qubits in semiconductor quantum dots are typically operated in
the lowest orbital to reduce interference from nearby states.
In hole-based systems, strong spin-orbit interaction couples spin and orbital degrees of freedom, significantly influencing the hole $g$-factor, a key parameter for qubit control. Precise knowledge of $g$-factors is therefore essential for reliable qubit operation.

Ge/SiGe heterostructures provide a favorable platform for hole quantum dots. Their low effective mass extends the wave function, easing gate design constraints~\cite{Hendrickx18}. 
In germanium, the g-tensor is strongly anisotropic~\cite{Hendrickxsweetspot24,Zhang_2021,Jirovec22,seidler2025spatialuniformitygtensorspinorbit}. The out-of-plane $g$-factor is typically \num{10} to \num{12} ~\cite{Hendrickxsweetspot24}, while the in-plane $g$-factor ranges from \num{0.01} to \num{1} ~\cite{Hendrickx20,Wang_Hopping_spins,Hendrickxsweetspot24,seidler2025spatialuniformitygtensorspinorbit} depending on magnetic field orientation. 
Current theory captures the general magnitude of these $g$-factors~\cite{brickson2024usinghighfidelitynumericalmodel} but fails to reproduce their angular dependence or precise values, likely due to orbital structure, confinement potential, or local strain. 

Probing how $g$-factors vary with hole number and orbital state is important for scaling qubit systems. 
Gate-defined quantum dots enable magnetospectroscopic measurements prior to qubit operation.
Coulomb blockade addition spectroscopy (CBAS) measures many-body energy differences between successive hole numbers~\cite{Tarucha_shell_96}, while pulsed excited-state spectroscopy (PESS) reveals excitation spectra at a fixed occupancy~\cite{Miller22,Kouwenhoven97_2}. Comparing these methods within the same dot allows separation of pure Zeeman contributions from orbital effects, helping to clarify discrepancies between methods.

In this study, we investigate a gate-defined quantum dot in a planar Ge/SiGe heterostructure using an integrated charge sensor. We measure energy spectra of the first few hole states with CBAS and PESS, extract Zeeman splittings, and compare $g$-factors obtained from addition and excitation spectra. By tuning the confinement potential with gate voltages, we also explore gate-induced modifications of the $g$-factor, demonstrating tunability relevant for all-electric qubit control.


\begin{figure*}
\includegraphics[width=\textwidth]{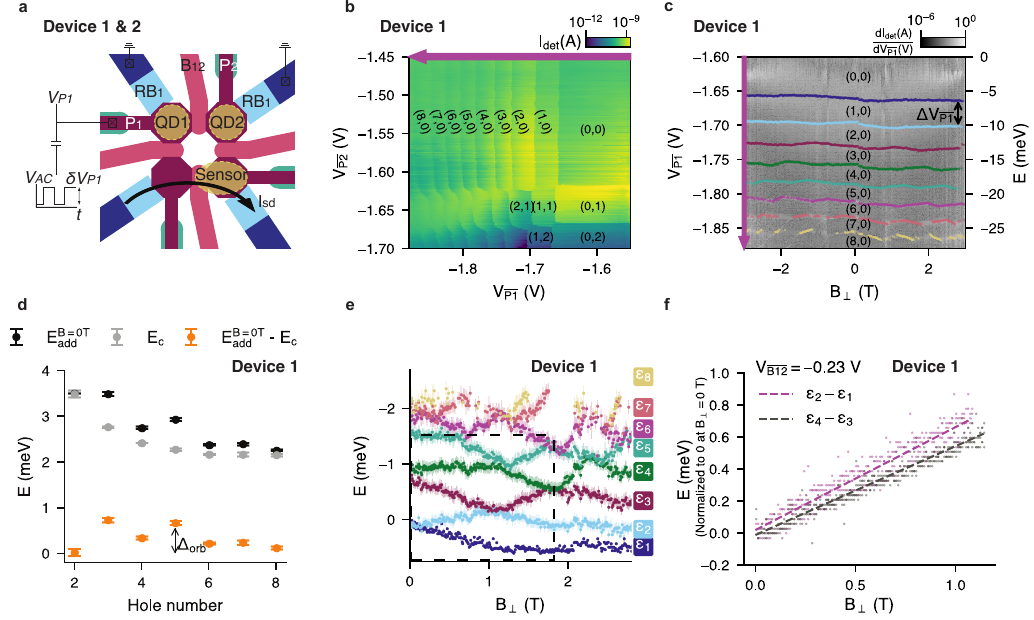}
\caption{\textbf{Operation of the double quantum dot (DQD) system and Coulomb blockade addition spectroscopy.} 
\textbf{(a)} Schematic of the device. The DQD is defined beneath plunger gates P$_1$ and P$_2$ and is monitored via a proximal charge sensor. Interdot coupling is tuned via the central barrier gate B$_{12}$, while coupling to the reservoirs is controlled by gates RB$_1$ and RB$_2$. The charge sensor signal is extracted from the differential current between source and drain, $I_{\mathrm{SD}}$, as indicated by the black arrow. 
\textbf{(b)} Charge stability diagram of the DQD as a function of virtual plunger gates $\vPone$ and $\vPtwo$. Charge configurations are labeled as $(N,M)$, denoting the hole occupation of each dot. 
\textbf{(c)} Magnetospectroscopy of Coulomb blockade addition energies, revealing spin and orbital structure. Charge states are again labeled as $(N,M)$. 
\textbf{(d)} Extracted addition energies (black), charging energies (gray), and single-particle energies (orange) as a function of total dot occupation. Error bars are smaller than the marker size. 
\textbf{(e)} Single-particle energy spectrum obtained by subtracting the charging energy $E_C$ from the addition energies in panel (d). The dashed box highlights the region analyzed in \figref{fig:fig2}b. 
\textbf{(f)} Linear fit to the spin-split energy levels used to extract the CBAS $g$-factor, based on the slope prior to the first level crossing.}
\label{fig:1}
\end{figure*}

Measurements were performed on two double quantum dot (DQD) devices, referred to as device 1 and device 2, to assess the reproducibility of the experimental findings. These devices, fabricated from the same wafer in different fabrication runs, were nominally identical. The dots were formed within the \SI{20}{\nano \meter} thick quantum well of a strained Ge/SiGe heterostructure~\cite{Hendrickxsweetspot24,Massai24}.
In each device, two quantum dots are located beneath plunger gates P$_1$ and P$_2$, with the tunable interdot coupling controlled by gate B$_{12}$ (\figref{fig:1}a). 
A nearby quantum dot charge sensor enables the detection of charge transitions. 
This technique allows us to probe the first few-hole states when the tunnel rate between the quantum dot and the reservoir is too low for direct transport measurements.
Using virtual gate voltages $\vPone$ and $\vPtwo$, we independently control the charge occupancy $(N,M)$ of each quantum dot (\figref{fig:1}b) (see \supmat{}~\ref{sec:methods} for details).
The devices are operated in a regime $(N,0)$ where quantum dot 2 (QD2) is fully depleted, and only quantum dot 1 (QD1) is populated. The occupancy of QD1 is tuned from $N=\num{0}$ to \num{8} holes along the arrow in \figref{fig:1}b. 


To reconstruct the energy spectrum of QD1, CBAS is performed on device 1 as a function of magnetic field $B_\perp$ applied perpendicular to the quantum well plane (see \figref{fig:1}c). The plunger gate voltage $\vPone$ is swept toward more negative voltages (arrow in \figref{fig:1}c) at fixed values of $B_\perp$, and the derivative $\diff I_\textnormal{det}/\diff V_{\overline{\textnormal{P1}}}$ is plotted normalized to its maximum value. The reverse sweep directions are also recorded but not shown. After each pair of sweeps,  $B_\perp$ is incremented, and the charge sensor is retuned to optimize sensitivity to the first charge transition. 
Charge transitions are identified by locating peaks in $\diff I_\textnormal{det}/\diff V_{\overline{\textnormal{P1}}}$, as a function of $\vPone$ for every $B_\perp$. 
These peak positions, marked by colored lines in \figref{fig:1}c, delineate the charge states indicated between them. 

The addition energy for each hole number $N$ is extracted from the spacing $\Delta \vPone$  between adjacent peaks in Fig.~\ref{fig:1}c using
$E_{\mathrm{add}}(N) =\mu_{N+1}-\mu_N=e\alpha_{\overline{\textnormal{P1}}}\Delta \vPone$,
with a gate lever arm $\alpha_{\overline{{P1}}} = \num{0.1}$ (see \supmat{} for details on~\ref{subsec:specextraction} spectrum and~\ref{subsec:lever} lever arm extraction).
\figref{fig:1}d shows these addition energies at $B_\perp=\SI{0}{\tesla}$ for the first eight holes (black points). Superimposed on a generally decreasing trend, we observe enhanced addition energies for adding the third and the fifth hole (hole numbers 2 and 4, respectively, in \figref{fig:1}d). This enhancement is a manifestation of the occupation of higher-energy orbital states in the dot, once the low-energy two-fold degenerate orbitals have been occupied by spin-up and spin-down holes.

To obtain an approximate magnetic-field-dependent single-particle spectrum $\epsilon_N(B_\perp)$ ($N=1,2,\ldots$) from the raw data (\figref{fig:1}c), we start from
\[ E_\textnormal{add}(N,B_\perp)=E_\textnormal{C}(N)+\Delta\epsilon_N(B_\perp),\]
where $E_\textnormal{C}(N)$ is the magnetic field independent charging energy for the dot with $N$ holes, and $\Delta\epsilon_N(B_\perp)=\epsilon_{N+1}(B_\perp)-\epsilon_N(B_\perp)$ is the magnetic field dependent spacing of excited states \cite{Ihn09}. We determine $E_\textnormal{C}(N)$ as the offset required to make the levels touch in one point, the crossing of different energy levels $\epsilon_{N}$, shown in \figref{fig:1}e. To define the energy reference, an offset is applied such that $\epsilon_1(B_\perp = \SI{0}{\tesla}) = 0$.

Before discussing the details of \figref{fig:1}e, we plot the extracted values of $E_\textnormal{C}(N)$ as gray points in \figref{fig:1}d.
The figure shows that the charging energy decreases monotonically with increased hole number, in contrast to the addition energy $E_\textnormal{add}(N)$. This effect is due to increased dot size, i.e. increased dot capacitance, driven by Coulomb interaction. 
This procedure leaves us with values $\Delta\epsilon_N(B_\perp)=E_\textnormal{add}(N,B_\perp)-E_\textnormal{C}(N)$, which are plotted for $B_\perp=\SI{0}{\tesla}$ as orange points in \figref{fig:1}d.  
These points reflect the alternating spin-filling of orbital levels mentioned before, with an enhanced $\Delta\epsilon_N$ for adding the third and fifth hole.

Returning to the spectrum $\epsilon_N(B_\perp)$ in \figref{fig:1}e, we see that  the two-fold degenerate levels for $N = 1, 2$ and $N = 3, 4$ split linearly for small $B_\perp < 1$~T. We interpret this splitting as an effective Zeeman splitting and determine the out-of-plane CBAS $g$-factor using
\[ g^{\epsilon_{N+1}-\epsilon_N} = \frac{1}{\mu_B}  \dv{(\Delta\epsilon_{N}(B_\perp))}{B_\perp},\] where the derivative is obtained from a linear fit to the Zeeman splitting (see \figref{fig:1}f), and $\mu_B$ is the Bohr magneton.
The extracted values are summarized in \tabref{tab:g}. The CBAS $g$-factor for the first spin pair, $g^{\epsilon_2 - \epsilon_1}$, agrees well by \SI{2}{\%} with values obtained from qubit spectroscopy~\cite{Hendrickxsweetspot24} via Pauli spin blockade readout. In contrast, the second spin pair, $g^{\epsilon_4 - \epsilon_3}$, exhibits a reduced $g$-factor, consistent with previous observations in Ref.~\cite{Miller22}. In this reference, the observed reduction of $g^{\epsilon_4-\epsilon_3}$ as compared to $g^{\epsilon_2-\epsilon_1}$ is attributed to mechanisms such as heavy-hole–light-hole mixing, leakage of the wave function into the SiGe barrier, or hole--hole interactions. The CBAS method reproduces the well-known strongly anisotropic $g$-tensor when the magnetic field orientation relative to the sample plane is changed (data shown in \supmat{} \ref{sec:tiltedfield}, Fig.~\ref{fig:fig_tilted}).

While CBAS provides an approximate single-particle level spectrum $\epsilon_N(B_\perp)$, it can be expected that this spectrum is not described by a single-particle model hamiltonian incorporating a fixed in-plane confinement potential, such as a Fock--Darwin model (see \supmat{}~\ref{sec:gBsquared}) or a similar model based on the Luttinger--Kohn hamiltonian.
For example, gate voltage sweeps alter the confinement potential, modifying the spectrum parametrically.
In addition, screening effects further modify the effective confinement potential, and exchange and correlation effects can be relevant. 
The CBAS procedure may also mask spin–orbit interaction effects that have been predicted to lead to avoided level crossings~\cite{Bulaev05_2}.

\begingroup

\setlength{\tabcolsep}{10pt} 
\renewcommand{\arraystretch}{1.7} 
\begin {table}
\caption {Extracted absolute $g$-factors with $3\sigma$-error bars from CBAS and PESS for both devices, where N is the transition hole number. Additionally, the orbital number investigated in the analysis are indicated.} \label{tab:g} 
\begin{center}
\begin{tabular}{ |c|c|c|c|c|}
\hline
\multicolumn{5}{|c|}{\textbf{Device 1}} \\
\hline
& method &  & $N$ & \#orbital \\
\hline
$g^{\epsilon_2 - \epsilon_1}$  & CBAS & \num{11.12 \pm 0.63}& \num{1},\num{2} & \num{1} \\
$g^{\epsilon_4 - \epsilon_3}$ & CBAS &\num{9.60 \pm 0.36}& \num{3},\num{4} & \num{2} \\
\hline
\multicolumn{5}{|c|}{\textbf{Device 2}} \\
\hline

$g^{\epsilon_2 - \epsilon_1}$ & CBAS &\num{11.25 \pm 0.33}&\num{1},\num{2} & \num{1} \\

\hline
\multicolumn{5}{|c|}{bare spin} \\
\hline
$g^{\uparrow_\text{o1} - \downarrow_\text{o1}}$ &PESS &\num{9.86\pm0.15}&\num{1} & \num{1}\\
$g^{T_{0} - T_{-}}$ &PESS&\num{10.25\pm0.63}&  \num{2}& \num{2}\\
\hline
\multicolumn{5}{|c|}{spin \& orbital} \\
\hline

$g^{\uparrow_\text{o1} - \downarrow_\text{o2}}$ &PESS &\num{13.71\pm0.33}&\num{1} & \num{1},\num{2}\\
$g^{S - T_{-} }$ &PESS&\num{14.25\pm0.33}&  \num{2}& \num{1},\num{2}\\

\hline
\end{tabular}
\end{center}
\end{table}
\endgroup

\begin{figure*}
    \includegraphics[width=\textwidth]{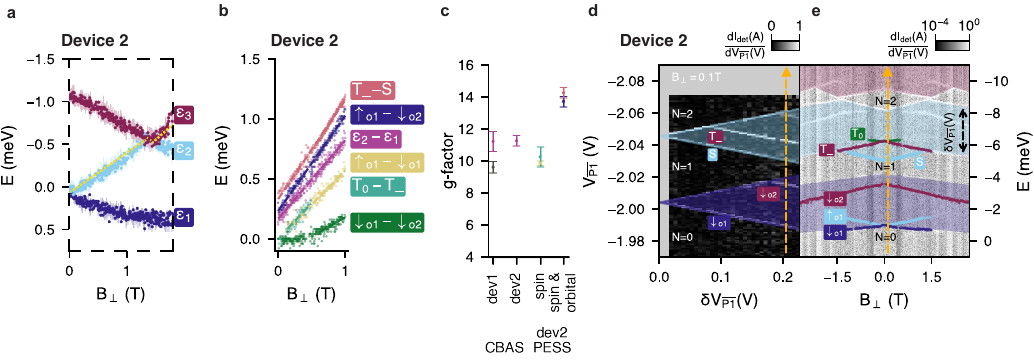}
    \caption{\textbf{Comparison of CBAS analysis to PESS.} 
\textbf{(a)} Energy spectrum obtained via CBAS. Ground-states of \textbf{d} from magnetospectroscopy are analyzed analogously. The yellow line marks the discontinuity at the intersection of $\epsilon_2$ and $\epsilon_3$.  
\textbf{(b)} Linear fits to the spin-split energy levels used to extract the CBAS and PESS $g$-factor. All data are shown in absolute values. For clarity, an offset is added to all traces except those corresponding to pure spin transitions ($\uparrow_\text{o1}-\downarrow_\text{o1}$ and $T_0-T_{-}$).
\textbf{(c)} CBAS and PESS $g$-factor for both devices with $3\sigma$-error bars.
\textbf{(d)} Derivative of the charge sensor current, $\diff I_\textnormal{det}/\diff V_{\overline{\textnormal{P1}}}$, plotted as a function of pulse amplitude $\delta V_{\mathrm{P1}}$ and DC gate voltage $\vPone$ at $B_\perp = 0$. Colored regions indicate where specific hole numbers can be dynamically loaded in response to the applied pulses. 
\textbf{(e)} $\diff I_\textnormal{det}/\diff V_{\overline{\textnormal{P1}}}$ as a function of $\vPone$ and $B_\perp$ at fixed $\delta \vPone$. The hole number $N$ is labeled, and regions sensitive to excited-state spectra are highlighted. Extracted energy levels are overlaid as colored lines. }
    \label{fig:fig2}
\end{figure*}

To directly probe the excited-state energies, we performed PESS measurements on device 2. To this end, we first convinced ourselves that
the addition spectrum of this device (see \figref{fig:fig2}a) shows qualitative agreement with device 1  (see \figref{fig:1}e). The dashed rectangle in both figures highlights the corresponding voltage range in the two devices, where we focus on the first three energy levels. The extracted CBAS $g$-factor $g^{\epsilon_2 - \epsilon_1}$ of device 2 is comparable to that of device 1 (see \tabref{tab:g} and \figref{fig:fig2}b). Additionally, $\Delta\epsilon_2$ is similar in both devices, ensuring consistency throughout the experiment. In both spectra  \figref{fig:1}e and \figref{fig:fig2}a, discontinuities are visible at the level crossings, highlighted with the dashed line in \figref{fig:fig2}a. It seems to be a multi-particle effect and is further discussed in \supmat{}~\ref{sec:STcrossing}.

In PESS following Ref. ~\cite{Liles18, Gaechter22}, rectangular virtualized AC pulses of amplitude $\delta \vPone$, frequency {$f=\SI{20}{\kilo\hertz}$}, and 50\% duty cycle are superimposed on the DC gate voltage $\vPone$.
This periodic modulation shifts the dot levels relative to the reservoir's Fermi level. The tunnel rates between the dot and the reservoir are measured by varying the pulse frequency, confirming that the tunnel rates are below \SI{1}{\kilo\hertz}, the minimum of the measurement range (see \supmat{} ~\ref{sec:Fig2supplement_tunnelrate}). An integration time of \SI{40}{\milli \second} is used to optimize the visibility of excited-states. 
Two branches appear as the pulse amplitude is increased corresponding to the edges of the two-level pulse (\figref{fig:fig2}d). 
For amplitudes $e\alpha_{\overline{\textnormal{P1}}}\delta \vPone$ exceeding an excitation energy at fixed hole number, additional lines appear parallel to the lower branch edge. Such excited states are observed both when filling the $N=1$ state in the transition region $N=0\leftrightarrow 1$ region (purple) and $N=2$ in the $N=1\leftrightarrow 2$ region (blue) in Fig.~\ref{fig:fig2}d.

Measuring PESS at a fixed pulse amplitude $\delta \vPone= \SI{0.2}{\volt}$ indicated by the dashed arrow in \figref{fig:fig2}d as a function of $\vPone$ and $B_\perp$ results in the raw spectrum shown in \figref{fig:fig2}e. 
In the $N=0\leftrightarrow1$ region, we observe the two-fold degenerate ground state at $B_\perp=\SI{0}{\tesla}$,  
that Zeeman-splits into two levels (purple, $\downarrow_\text{o1}$ and blue $\uparrow_\text{o1})$ at finite $B_\perp$ giving rise to the pure spin PESS $g$-factor $g^{\uparrow_\text{o1}-\downarrow_\text{o1}}$ shown in \tabref{tab:g}. The first excited orbital state (red $\downarrow_\text{o2}$) runs nearly parallel to the ground state, suggesting that it shares the same spin projection. However, its Zeeman splitting is not visible in our measurements, likely due to insufficient difference in tunnel rates. While the excited state $\downarrow_\text{o2}$ changes the tunnel rate sufficiently to produce a contrast in the averaged measurement, its Zeeman split partner $\uparrow_\text{o2}$ does not alter the tunneling dynamics sufficiently to be detected. 
In the $N=1\leftrightarrow 2$ region, the ground state (singlet state, blue $S$) is non-degenerate with total spin angular momentum $0$. The observed triplet states $T_-$ and $T_0$ involve a second excited orbital state, but are energetically lowered by exchange, placing them well below the single-particle orbital splitting observed for $N=0\leftrightarrow1$ ($\Delta \text{orb}_{N_{0\leftrightarrow1}}^{B_\perp=\SI{0}{\tesla}} = \SI{2.56}{meV}$, $\Delta \text{orb}_{N_{1\leftrightarrow2}}^{B_\perp=\SI{0}{\tesla}} = \SI{1.07}{meV}$). 
Additionally, the confinement potential is increasingly screened when adding more holes, which also reduces this splitting.
The triplet states Zeeman-split linearly into $T_-$ and $T_0$ at finite $B_\perp$ giving rise to the PESS $g$-factor $g^{T_0-T_-}$ in \tabref{tab:g}. An additional excited state at higher energy is also observed ~\cite{Ellenberg2006}, which is beyond our interest in this paper.
At the onset of the $N=2\leftrightarrow 3$ region, we identify the ground-state with the total angular momentum of the $N=1$ ground-state. Overall, the data suggest that states fill in a sequence of alternating sign of angular momentum $z$-projection.

Figure \ref{fig:fig2}b shows a direct comparison of the energy level differences from CBAS and PESS measurements used to determine the CBAS and PESS $g$-factors in \tabref{tab:g}.
The excited-state pairs highlighted in Fig.~\ref{fig:fig2}e allow extraction of $g$-factors using the slope-based method of the linear splitting with magnetic field as described earlier. The values, shown in Fig.~\ref{fig:fig2}c (here we show $3\sigma$-error bars) and summarized in Table~\ref{tab:g}, reveal that $g^{\uparrow_\text{o1}-\downarrow_\text{o1}}$ (yellow) is reasonably close to $g^{T_0 - T_-}$ (turquoise) within experimental error, where both are pure spin $g$-factors. This contrasts with the CBAS $g$-factor (purple) which is significantly larger, highlighting the fact that the chosen spectroscopy method can have an important influence on the $g$-factor. 

Interestingly, the reduction of the CBAS $g^{\epsilon_4 - \epsilon_3}$ as compared to $g^{\epsilon_2 - \epsilon_1}$, discussed before, is not observed when comparing the corresponding PESS $g^{T_0-T_-}$ to $g^{\uparrow_\text{o1}-\downarrow_\text{o1}}$.

To investigate the orbital contribution to the $g$-factors within the same charge configuration, we compute the energy difference between the first excited orbital state (red, $\downarrow_\text{o2}$) and the ground orbital state (purple, $\downarrow_\text{o1}$), and plot this difference in green in Fig.~\ref{fig:fig2}b. This difference remains approximately constant up to \SI{0.5}{\tesla}, after which it begins to increase. By comparing the pure Zeeman level splitting ($\downarrow_\text{o1}-\uparrow_\text{o1}$) in the range of zero to one Tesla to the change in orbital level splitting between ground and first excited state ($\downarrow_\text{o1}-\downarrow_\text{o2}$), we estimate that changes in orbital wave function can contribute up to 10\% to the apparent Zeeman splitting measured between spin states with distinct orbital wave functions, such as ${S-T_{-}}$. The nonlinearities caused by such orbital effects appear as systematic deviations of linear fits from the data (see \supmat{} \figref{fig:residual_PESS}e). Beyond this experimental evidence for orbital influences on the $g$-factor determination, we confirm a similar order of magnitude from simple theoretical considerations (\supmat{} \ref{sec:sup_EnergyComparison}). 
Based on these findings, when orbital states are included in the PESS $g$-factor extraction, the orbital contribution appears to add to the bare spin component resulting in an overall increase in the PESS $g$-factor for $g^{\uparrow_\text{o1}-\downarrow_\text{o2}}$ ($N=0\leftrightarrow1$) and $g^{S - T_-}$ ($N=1\leftrightarrow2$), as summarized in Table \ref{tab:g}. Similar to CBAS, PESS can also be used to measure the anisotropic $g$-tensor (see \supmat{} \ref{sec:tiltedfield}, Fig.~\ref{fig:fig9_tilted}).

\begin{figure*}[ht!]
    \includegraphics[width=\textwidth]{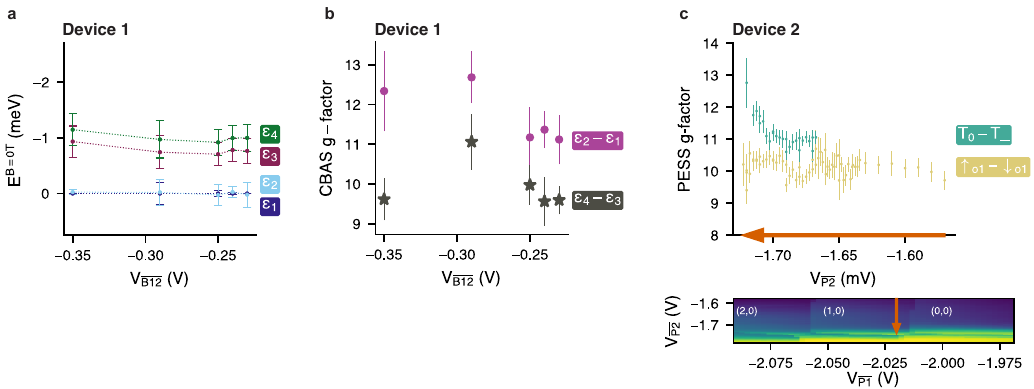}
    \caption{\textbf{Voltage-tunable $g$-factor.} 
\textbf{(a)} Single-particle energies at $B_\perp=\SI{0}{\tesla}$ as a function of quantum dot occupation for different values of $V_{\overline{B{12}}}$. 
\textbf{(b)} CBAS $g$-factor extracted via CBAS as a function of the interdot barrier voltage $V_{\overline{B{12}}}$. 
\textbf{(c)} PESS $g$-factor as a function of the plunger gate voltage applied to the neighboring quantum dot. The lower plot shows the corresponding charge stability diagram, with arrows indicating the direction of plunger gate tuning. The charge configuration is labeled as $(N,M)$.} 
\label{fig:fig3}
\end{figure*}
Having established the key differences between the two methods for $g$-factor determination, we now examine the voltage tunability of the $g$-factor. Such a tunability would greatly enable gate-driven qubit operation~\cite{Abadillo23,Maier13}.  
By varying the voltages applied to adjacent gate electrodes, we track the evolution of the $g$-factor using both CBAS and PESS measurement techniques.
For device 1, we investigate the tunability of the CBAS $g$-factor by varying the interdot barrier gate voltage $V_{\overline{B{12}}}$ from \SI{-0.23}{\volt} to \SI{-0.35}{\volt} (the charge stability diagrams can be found in \supmat{}~\ref{sec:CBAS_fulldata}). This voltage range is chosen to avoid hysteretic gate response inherent to the device~\cite{Massai24}.
Figure~\ref{fig:fig3}a shows the approximate single-particle energies at $B_\perp=\SI{0}{\tesla}$ for the first four levels. They remain nearly unchanged within experimental uncertainty ($3\sigma$-error bars). We interpret this as an indication that the confinement potential is primarily shifted in space without significantly altering its characteristic shape.

Figure~\ref{fig:fig3}b shows the CBAS $g$-factors for the first two Zeeman-split pairs. Changes in $g$-factor beyond experimental uncertainty are observed at the level of 15\%.
The extracted CBAS $g$-factor $g^{\epsilon_2 - \epsilon_1}$ exhibits a more consistent increase for increasingly negative $V_{\overline{B{12}}}$ compared to $g^{\epsilon_4 - \epsilon_3}$.
Unfortunately, changing the barrier gate $V_{\overline{B{12}}}$ leads to a significant cross-talk reducing the tunnel coupling of the dot to the lead. This effect prohibited measurements of the $g$-factor using PESS, for which the tunnel rate must remain within the operational bandwidth of our electronics ($\sim 10^3$–$10^7$~Hz). To still enable PESS $g$-factor measurements, we employ the virtual plunger gate $V_{\overline{P_{2}}}$ instead. 
By sweeping the plunger gate of the neighboring dot toward the interdot regime, as indicated by the arrows in the bottom of \figref{fig:fig3}c, the quantum dot wavefunction gradually shifts from beneath $\vPone$ towards $\vPtwo$.
The extracted pure spin PESS $g$-factors are shown in Fig.~\ref{fig:fig3}c (data for the second dot in \supmat{}~\ref{sec:tiltedfield}).
While $g^{\uparrow_\text{o1}-\downarrow_\text{o1}}$ remains constant within experimental uncertainty, $g^{T_0-T_{-}}$  varies approximately by $20\%$ as the wavefunction delocalizes across the barrier and subsequently relocalizes in the second dot. This sensitivity likely arises from changes in the local potential and strain as the hole’s position shifts relative to the interdot barrier, consistent with previous reports on electrostatic and strain-induced $g$-factor variability~\cite{seidler2025spatialuniformitygtensorspinorbit,martinez2025variabilityholespinqubits,Abadillo23}.

In conclusion, we have identified an experimental method to determine the pure Zeeman $g$-factor in p-type germanium quantum dots and to distinguish it from orbital contributions arising from the strong spin-orbit coupling in this material. The significant influence of orbital effects at the level of 10\% of the bare Zeeman splitting in the magnetic field range between 0 and \SI{1}{T} adds to explanations for the variations in the extracted $g$-factors found between different methods (CBAS vs PESS), across orbital states, and for different hole numbers. Additional influences are expected from changes in gate-induced changes to the confinement potential and from screening-, exchange- and correlation-effects relevant in dots of more than one hole. These findings highlight the complexity of an accurate characterization of $g$-factors in hole quantum dots and indicate that comparisons between $g$-factors obtained in the community need to be made with great care. With these insights in mind we have explored the opportunity to gate-control the $g$-factors. A tunability of up to 15\% was found by shifting the quantum dot states spatially, which may indicate an opportunity for all-electric qubit manipulation.

\begin{acknowledgments}
The authors thank Michael Stiefel and all the Cleanroom Operations Team of the Binnig and Rohrer Nanotechnology Center (BRNC) for their help and support. The authors thank Daniel Loss, Jelena Klinovaja, Christoph Adelsberger, Dmitry Miserev and Atreyee Basu for fruitful discussions.
\paragraph*{\bf Funding}
This research was funded in part by {NCCR SPIN}, a National Centre of Competence in Research, funded by the Swiss National Science Foundation (grants  \mbox{51NF40-180604} and \mbox{51NF40-225153}) and by the Swiss National Science Foundation (grant \mbox{200021-188752}).
\paragraph*{\bf Author contributions}

L.S. performed the experiments and data analysis; F.J.S. and M.M. fabricated the device 
and N.W.H designed the gate layout with contributions from L.S., I.S., L.M., A.O., K.T. and E.G.K.; S.P. with help from L.S., I.S. and P.H.C. contributed to the development of the experimental setup; S.W.B. provided the heterostructures; G.S., I.S., P.H.C. developed the measurement software;
L.S. wrote the manuscript with contributions from I.S., P.H.C., T.I., A.F., and input from all authors; P.H.C., A.F. and T.I.  discussed the results and supervised the project.

\paragraph*{\bf Competing interests}
The authors declare no competing interests.
\paragraph*{\bf Data availability}
The data underlying this study will be made available in a Zenodo repository.
\end{acknowledgments}

\bibliographystyle{apsrev4-1-title} 
\bibliography{references.bib}


\makeatletter

\renewcommand\thesubsection{\mbox{S\arabic{section}}}

\renewcommand\thesection{\mbox{S\arabic{section}}}
\makeatother

\newcounter{supfigure} \setcounter{supfigure}{0} 
\makeatletter
\renewcommand\thefigure{\mbox{S\arabic{supfigure}}}
\makeatother

\newcounter{suptable} \setcounter{suptable}{0} 
\makeatletter
\renewcommand\thetable{\mbox{S\arabic{suptable}}}
\makeatother

\newenvironment{supfigure}[1][]{\begin{figure}[#1]\addtocounter{supfigure}{1}}{\end{figure}\ignorespacesafterend}
\newenvironment{supfigure*}[1][]{\begin{figure*}[#1]\addtocounter{supfigure}{1}}{\end{figure*}\ignorespacesafterend}
\newenvironment{suptable}[1][]{\begin{table}[#1]\addtocounter{suptable}{1}}{\end{table}\ignorespacesafterend}
\newenvironment{suptable*}[1][]{\begin{table*}[#1]\addtocounter{suptable}{1}}{\end{table*}\ignorespacesafterend}

\makeatletter
\renewcommand\theequation{S\arabic{equation}}
\renewcommand\theHequation{S\arabic{equation}} 
\makeatother
\setcounter{equation}{0}

\onecolumngrid 
\clearpage
{\centering
\large\textbf
{Supplementary information for: \\ \mytitle} \\ \rule{0pt}{12pt}
}
\twocolumngrid

\newcommand{\RefFigResonant}{\figref{fig:resonant}} 
\newcommand{\RefFigDipersivePhi}{\figref{fig:dispersivephi}} 
\newcommand{\RefFigDipersivePhotonNumber}{\figref{fig:dispersivephotonnumber}} 

\newcommand{\RefExcited}{\secref{sec:excited_Spect}} 
\newcommand{\RefFigthree}{\figref{fig:fig3}} 
\newcommand{\RefFigtwo}{\figref{fig:fig2}} 


\renewcommand{\thefigure}{S\arabic{figure}}
\setcounter{figure}{0}

\section{\label{sec:methods} Methods}

\subsection*{\label{subsec:fab} Device fabrication}

The devices are based on a Ge/SiGe heterostructure with a \SI{20}{\nano \meter}-thick strained Ge quantum well located \SI{48}{\nano \meter} below the wafer surface. The heterostructure, with a Si$_{0.2}$Ge$_{0.8}$ barrier composition, is grown using an industrial reduced-pressure chemical vapor deposition (RP-CVD) process.
Ohmic contacts to the quantum well are formed by platinum diffusion at \SI{300}{\celsius}, resulting in low-resistance Pt-silicide contacts. The gate layout is fabricated in two layers. In the first layer, electrostatic gates are defined via electron-beam lithography and lift-off of a \SI{20}{\nano \meter} Ti/Pd metal stack. A \SI{7}{\nano \meter}-thick SiO$_2$ dielectric, deposited by plasma-enhanced atomic layer deposition (PE-ALD), electrically isolates the gate layers.
This study includes measurements from two devices fabricated in separate runs on the same wafer, demonstrating reproducibility across fabrication batches.

\subsection*{\label{subsec:setup} Experimental setup}

All measurements were performed in a Bluefors LD400 dilution refrigerator with a base temperature of $T_{mxc}$ $\approx$ \SI{15}{\milli \kelvin}. The device, mounted on a QDevil QBoard circuit board, was controlled via static gate voltages applied using a QDevil QDAC. DC lines were filtered at the mixing chamber stage using QDevil QFilters, guided through low-noise looms.
Sensor conductance was measured using two Basel Precision Instruments (BasPI) SP983c IV converters (gain: $10^9$, output low-pass filter: \SI{300}{\hertz}), with a source-drain bias of $V_{SD} = $\SI{200}{\micro \volt} for device 1 and \SI{500}{\micro \volt} for device 2. The output was recorded using a Keysight 34461A digital multimeter.
Magnetospectroscopy was conducted using an American Magnetics three-axis vector magnet, capable of applying fields up to (1, 1, 6)~T in the (x, y, z) directions, with high-stability current sources on all axes. The device was centered with respect to the z-axis solenoid. Minor angular misalignments in tilted field measurements can be caused by imperfect planar mounting. Magnetic hysteresis effects, typically a few mT, were negligible in our field range.
Pulsed excited-state spectroscopy was performed on a second, nominally identical device in a separate Bluefors LD400 fridge (base temperature $\approx$~13~mK) using the same electronics. For pulsing, bias tees with a \SI{1.1}{\milli s} time constant were employed. The high-frequency lines had an attenuation of \SI{22}{dB}.

\subsection*{Virtual gate matrices}
\label{subsec:virt}
To mitigate capacitive crosstalk between the different electrostatic gates and the quantum dots, the following virtual gates~\cite{Hensgens17} are used for the device 1:
\newcommand\scalemath[2]{\scalebox{#1}{\mbox{\ensuremath{\displaystyle #2}}}}

\[
\scalemath{0.9}{
\begin{pmatrix}
V_{\mathrm{P{1}}}\\
V_{\mathrm{P{2}}}\\
V_{\mathrm{P{3}}}\\
V_{\mathrm{B{12}}}\\
V_{\mathrm{CP{2}}}\\
V_{\mathrm{RB{2}}}
\end{pmatrix}
=
\begin{pmatrix}
1 & -0.220 & -0.092 & 0 & 0 & 0\\
0 & 1 & 0 & 0 & 0 & 0\\
-0.105 & -0.239 & 1 & -0.381 & 0 & 0\\
0 & 0 & 0 & 1 & 0 & 0 \\
-0.4 & 0 & 0 & 0 & 1 & 0 \\
-0.4 & 0 & 0 & 0 & 0 & 1 
\end{pmatrix}
\begin{pmatrix}
V_{\overline{\mathrm{P{1}}}}\\
V_{\overline{\mathrm{P{2}}}}\\
V_{\overline{\mathrm{P{3}}}}\\
V_{\overline{\mathrm{B{12}}}}\\
V_{\overline{\mathrm{CP{2}}}}\\
V_{\overline{\mathrm{RB{2}}}}
\end{pmatrix}
}
\]
$V_{\mathrm{Gi}}$ are the real and $V_{\overline{\mathrm{Gi}}}$ the virtual gate voltage, which leaves the chemical potentials of the nearby quantum dots unchanged. 

For the device 2 the virtual gatematrix is:

\[
\scalemath{0.7}{
\begin{pmatrix}
V_{\mathrm{P{1}}}\\
V_{\mathrm{P{2}}}\\
V_{\mathrm{P{3}}}\\
V_{\mathrm{B{12}}}\\
V_{\mathrm{B{23}}}\\
V_{\mathrm{B{41}}}\\
V_{\mathrm{RB{1}}}\\
V_{\mathrm{RB{2}}}
\end{pmatrix}
=
\begin{pmatrix}
1 & -0.239 & 0 & 0 & 0 & 0 & 0 & 0\\
-0.345 & 1 & 0 & 0 & 0& 0 & 0 & 0\\
-0.14 & -0.19 & 1 & -0.3 & -0.66 &-0.47 &-0.1 & -0.055\\
0 & 0 & 0 & 1 & 0& 0 & 0 & 0\\
0 & 0 & 0 & 0 & 1& 0 & 0 & 0\\
0 & 0 & 0 & 0 & 0& 1 & 0 & 0\\
0 & 0 & 0 & 0 & 0& 0 & 1 & 0\\
0 & 0 & 0 & 0 & 0& 0 & 0 & 1\\
\end{pmatrix}
\begin{pmatrix}
V_{\overline{\mathrm{P{1}}}}\\
V_{\overline{\mathrm{P{2}}}}\\
V_{\overline{\mathrm{P{3}}}}\\
V_{\overline{\mathrm{B{12}}}}\\
V_{\overline{\mathrm{B{23}}}}\\
V_{\overline{\mathrm{B{41}}}}\\
V_{\overline{\mathrm{RB{1}}}}\\
V_{\overline{\mathrm{RB{2}}}}
\end{pmatrix}
}
\]
\section{\label{subsec:specextraction} Single-particle spectrum extraction}
The magnetospectroscopy data is acquired by stepping the magnetic field from \SI{-3}{\tesla} to \SI{3}{\tesla}, while sweeping $\vPone$ from positive to negative and back. The derivative $\diff I_\textnormal{det}/\diff V_{\overline{\textnormal{P1}}}$ is normalized to its maximum value, and peak positions are extracted. To suppress effects from charge rearrangements, drift, and other slow variations, the differences of two adjacent peaks are calculated. 
Both forward and backward sweeps are included to capture potential hysteretic behavior, and the final dataset represents the mean of the two. Error bars shown in \figref{fig:1}e correspond to the standard deviation of this mean. Since the energy differences are symmetric with respect to magnetic field, the data are plotted as a function of the absolute value of $B_\perp$, resulting in two data points per magnetic field value.
While for CBAS the sensor is retuned for each $B_\perp$ value, this is not done for PESS since it remained sufficient sensitivity throughout the magnetic field sweep.

\section{\label{subsec:lever} Lever arm extraction}

\begin{figure*}
    \includegraphics[width=\textwidth]{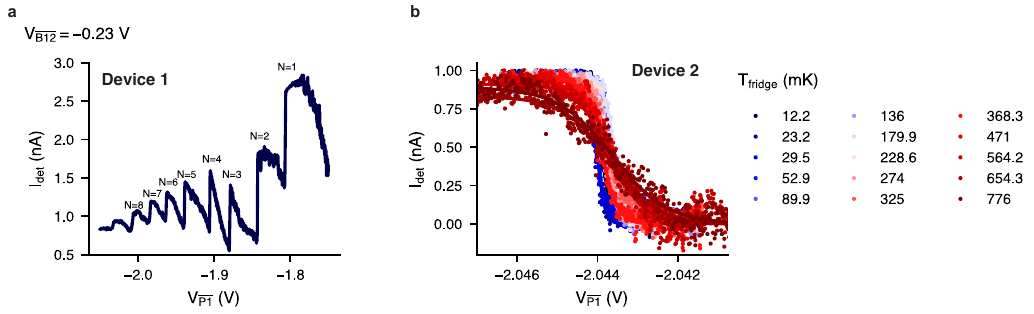}
    \caption{\textbf{Extraction of the lever arm from temperature-broadened transitions.} 
    \textbf{(a)} Example dataset showing eight charge transitions measured via the sensor current $I_{\mathrm{SD}}$ for device 1 at $V_{\overline{\mathrm{B{12}}}}$ is \SI{-0.23}{\volt}. 
    \textbf{(b)} Zoom-in of the first transition, measured on device 2. The sensor current is normalized, and the transition is fit using a Fermi–Dirac distribution at various temperatures, illustrating the thermal broadening.}
    \label{fig:lever_arm_Tsweep}
\end{figure*} 
The lever arms of the virtual gates $\vPone$ and $\vPtwo$ were extracted using temperature broadening of Coulomb peaks as a function of the fridge temperature. To this end, Coulomb blockade transitions were measured at \( B_{\perp} = \SI{0}{\tesla} \) in both sweep directions and repeated multiple times (\figref{fig:lever_arm_Tsweep}). 
Higher occupancies have higher uncertainty due to reduced measurement sensitivity.
Each transition was fitted using a Fermi-Dirac distribution after background subtraction, normalization, and voltage offset correction to center the transition around zero:

\begin{equation}
\label{eq:Fermi}
f(V)=\frac{A}{e^{\frac{\alpha(V-V_0)}{k_B\text{T}_\text{e} }}+1}
\end{equation}

\begin{figure*}
    \includegraphics[width=\textwidth]{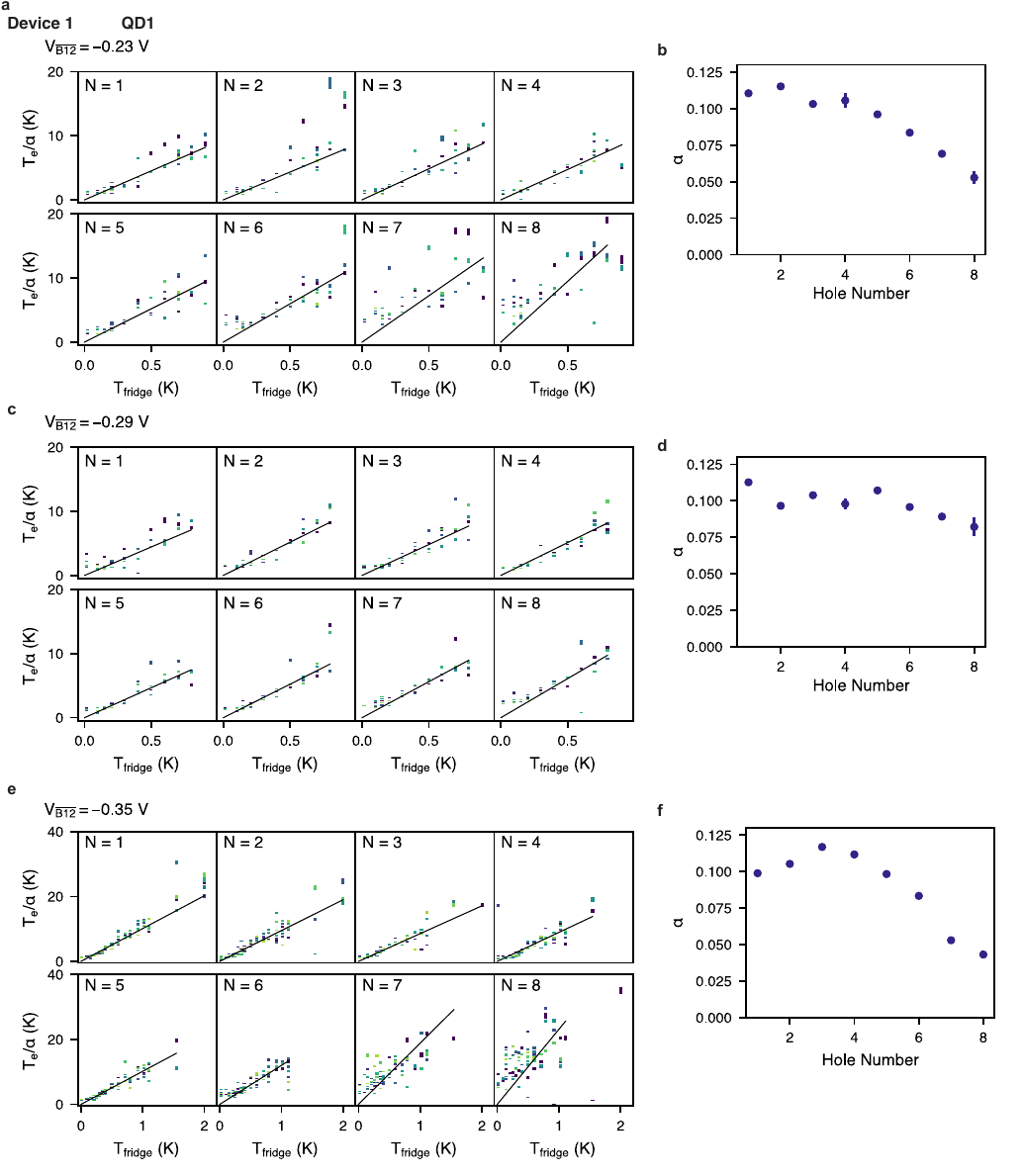}
    \caption{\textbf{Extraction of the lever arm from temperature-broadened transitions for device 1.} 
    \textbf{(a, c, e)} Width of the Fermi–Dirac distribution extracted from the data in \figref{fig:lever_arm_Tsweep}, plotted as a function of fridge temperature for the $N$th transition, and fitted for different values of $V_{\overline{\mathrm{B{12}}}}$. 
    \textbf{(b, d, f)} Resulting lever arms plotted as a function of hole number $N$ for the corresponding $V_{\overline{\mathrm{B{12}}}}$ values.}
    \label{fig:lever_arm_extr}
\end{figure*}

\begin{figure*}
    \includegraphics[width=\textwidth]{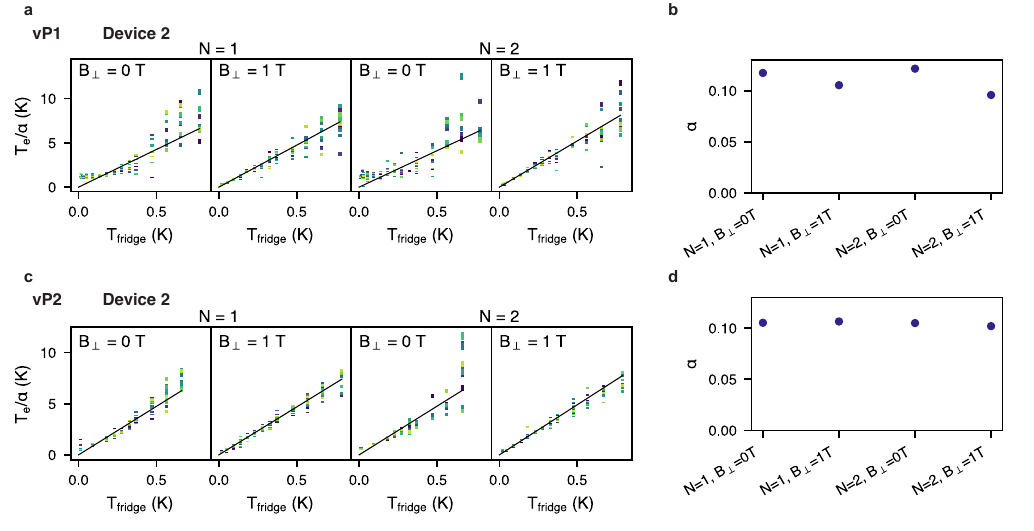}
    \caption{\textbf{Extraction of the lever arm from temperature-broadened transitions for device 2.} 
    \textbf{(a, c)} Width of the Fermi–Dirac distribution extracted from the data in \figref{fig:lever_arm_Tsweep}, plotted as a function of fridge temperature for the $N$th transition, and fitted for different out-of-plane magnetic fields $B_\perp$. 
    \textbf{(b, d)} Resulting lever arms plotted as a function of hole number $N$ and magnetic field $B_\perp$.}
    \label{fig:lever_arm_extr_PESS}
\end{figure*}

Here, the amplitude $A$ and the voltage offset $V_0$ are known parameters, which were allowed to vary during the fitting, and the primary fitting parameter is the ratio \( \frac{T_\text{e}}{\alpha} \), where $T_\text{e}$ is the base electron temperature and ${\alpha}$ is the lever arm. Fits were performed across multiple datasets, and only those with \( R^2 > 0.9 \) were retained for further analysis. The extracted \( T_\text{e} / \alpha \) values were plotted against the stabilized fridge temperature \( T_\text{fridge} \) (\figref{fig:lever_arm_extr}a,c,e and \figref{fig:lever_arm_extr_PESS}a,c). Temperature readings were taken after a \SI{15}{min} stabilization period. The heater was mounted close to the device, while the temperature sensor was positioned on the opposite side, potentially introducing a thermal gradient. However, both components were thermally contacted to the probe’s metal body, ensuring good thermal contact and minimizing systematic errors in \( T_\text{fridge} \).

At low temperatures, the broadening remains constant up to approximately \SI{200}{\milli\kelvin}, indicating a base electron temperature \( T_e \). Beyond this point, the broadening increases linearly with \( T_\text{fridge} \). A linear fit with zero intercept yields the lever arm \( \alpha \) as the slope. Higher temperatures lead to increased data scatter, reducing fit accuracy.
The extracted lever arms for different hole numbers and for the second device (including out-of-plane magnetic field data) are shown in \figref{fig:lever_arm_extr}b,d,f and \figref{fig:lever_arm_extr_PESS}b,d. Uncertainties represent the standard deviation of the fit and are in some cases smaller than the marker size. However, the spread in \( T_\text{e} / \alpha \) resulting in a minimum uncertainty of \SI{10}{\percent}, which propagates into the lever arm estimation.

Across all datasets, the lever arm values consistently converge around \( \alpha \approx 0.1 \).


\begin{figure*}
    \includegraphics[width=\textwidth]{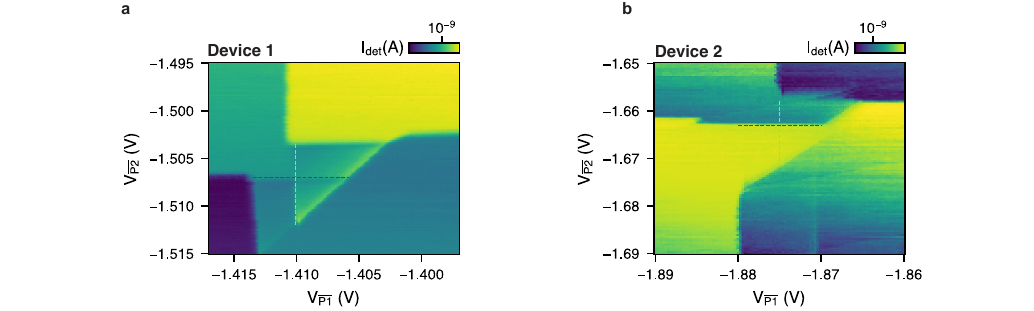}
    \caption{\textbf{Extraction of the lever arm from bias triangles.} 
    Bias triangles are measured by applying a source–drain bias of \SI{1}{\milli\volt} to the double quantum dot. Dashed lines indicate the voltage span used for lever arm extraction. 
    \textbf{(a)} Measurement of device 1.  \textbf{(b)}  Measurement of device 2.}
    \label{fig:lever_arm_extr_transport}
\end{figure*}

To validate the lever arm extraction, we employed an independent method based on transport measurements. In this approach, a finite bias was applied across the double quantum dot in addition to the sensor bias. Bias triangles were recorded at the $(1,0)-(0,1)$ interdot transition. The lever arm was then determined by dividing the applied bias voltage by the extent of the bias triangles along the gate voltage axis, as indicated by the dashed lines in \figref{fig:lever_arm_extr_transport}.
For device 1, this analysis yielded lever arms of $\alpha_{V_{\overline{\mathrm{P1}}}} = 0.113$ and $\alpha_{V_{\overline{\mathrm{P2}}}} = 0.103$, while for device 2, we obtained $\alpha_{V_{\overline{\mathrm{P1}}}} = 0.100$ and $\alpha_{V_{\overline{\mathrm{P2}}}} = 0.077$. Based on the consistency of these values, we adopt a representative lever arm of \num{0.1} for both virtual plunger gates ($\vPone$ and $\vPtwo$) across both devices throughout this work.

\section{\label{sec:gBsquared} Limitations of the Fock-Darwin model}
For the CBAS within the isotropic Fock-Darwin model, energy spectrum is given by~\cite{Ihn09}:
\[
E_{n,l}=\hbar\Omega(2n+|l|+1)-\frac{1}{2}\hbar\omega_cl\pm \frac{1}{2}g_N^*\mu_B B
\]
where $\Omega=(\omega_N^2+(\omega_c/2)^2)^{(1/2)}$, $\omega_N$ characterizes the confinement potential of the dot with $N$ holes, $g^*_N$ is the effective $g$-factor for hole number $N$ and $\omega_c=eB/m^*$. To analyze the influence of the orbital state on the slope of the energy-level difference, we calculate it explicitly for small magnetic fields ($B<\SI{0.2}{\tesla}$), where higher-order terms in $B^{4}$ can be neglected. As an example, we consider the first two levels, $E_{0,0}$ and $E_{0,1}$, both with spin $+1/2$. The difference is given by:
\[
\begin{split}
E_{0,1}-E_{0,0}=\hbar\underbrace{(\omega_1-\omega_0)}_{<0}+\frac{1}{2}\left( \frac{eB}{2m}\right)^2 \underbrace{\left( \frac{2}{\omega_1}-\frac{1}{\omega_0}\right)}_{>0}+\\
\frac{\hbar \omega_c}{2}\mp\frac{1}{2}\mu_B B (g^*_1-g^*_0)
\end{split}
\]

\begin{figure}
    \centering
    \includegraphics[width=\linewidth]{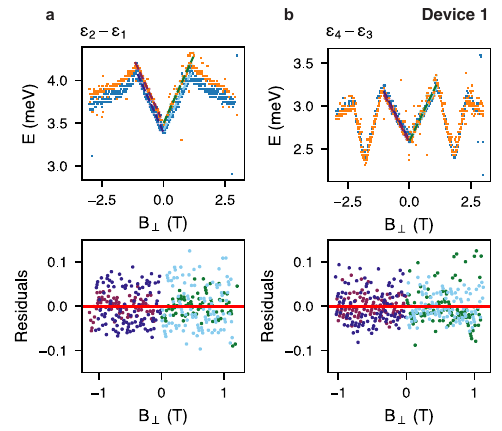}
    \caption{\textbf{Residual analysis from linear fits to CBAS data of device 1} Residuals are extracted from linear fits to the difference of levels within the CBAS measurements shown in \figref{fig:1}e. Forward and backward magnetic field sweeps and voltage sweeps are analyzed separately to better resolve potential trends.
    \textbf{(a)}  $g$-factor extraction for the ground states of $N=1,2$, as obtained from CBAS measurements ($g^{\epsilon_2-\epsilon_1}$). The residuals show no significant trend.
    \textbf{(b)} $g$-factor extraction for the ground states of $N=3,4$, as obtained from CBAS measurements ($g^{\epsilon_4-\epsilon_3}$). The residuals show no significant trend.
    }
    \label{fig:residual_dev1}
\end{figure}

\begin{figure*}
    \centering
    \includegraphics[width=\textwidth]{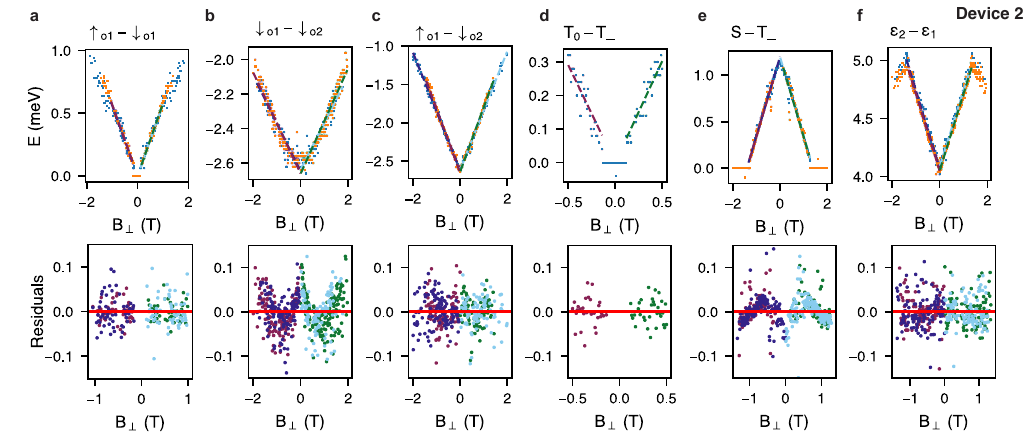}
    \caption{\textbf{Residual analysis from linear fits to PESS data of device 2} Residuals are extracted from linear fits to the difference of levels within the PESS measurements shown in \figref{fig:fig2}d. Forward and backward magnetic field sweeps and voltage sweeps are analyzed separately to better resolve potential trends.
    \textbf{(a)}  Extraction of the pure spin $g$-factor for the $N=1$ ground state ($g^{\uparrow_\text{o1}-\downarrow_\text{o1}}$). The residuals show no significant trend.
    \textbf{(b)} linear fit extraction for $N=1$ including the first excited orbital state(${\downarrow_\text{o1}-\downarrow_\text{o2}}$). The residuals show a systematic deviation, suggesting an orbital contribution.
    \textbf{(b)} $g$-factor extraction for $N=1$ including the first excited orbital state($g^{\uparrow_\text{o1}-\downarrow_\text{o2}}$). The residuals show a large spread, but no systematic deviation.
    \textbf{(c)} Pure spin $g$-factor extraction for the $N=2$ transition ($g^{T_0-T_{-}}$). The residuals show no significant trend.
    \textbf{(d)} $g$-factor extraction for $N=2$ including the lower orbital state ($g^{S-T_{-}}$). The residuals exhibit a systematic deviation, suggesting an additional orbital contribution that introduces a $B^2$ dependence. 
    \textbf{(e)} $g$-factor extraction for the ground states of $N=1,2$, as obtained from CBAS measurements ($g^{\epsilon_2-\epsilon_1}$). The residuals show no significant trend.
    }
    \label{fig:residual_PESS}
\end{figure*}

This analysis reveals that the change in the confinement potential by hole number $N$ increase introduces a quadratic dependence on the magnetic field. While this effect may not be directly visible in the raw data, it becomes apparent in the residuals of the linear fits, as shown in \figref{fig:residual_PESS} and \figref{fig:residual_dev1}. Notably, a clear deviation from linearity is observed only in the extraction of $g^{S-T_{-}}$, indicating that the inclusion of different orbital states seems to introduce an additional $B^2$ effect. For the other extracted $g$-factors, the residuals remain featureless, suggesting that changes in the confinement potential are negligible in those cases even in CBAS $g^{\epsilon_2-\epsilon_1}$ and $g^{\epsilon_4-\epsilon_3}$, where it should be most evident \figref{fig:residual_dev1}. This shows the limitations of the Fock-Darwin model.

Assuming an anisotropic Fock-Darwin model for including an ellipsoidal dot the energy spectrum is given by~\cite{Madhav94}:
\[
E_{n_x, n_y} = (n_x+\frac{1}{2})\hbar\omega_1+(n_y+\frac{1}{2})\hbar\omega_2\pm\frac{1}{2}g_N^*\mu_BB
\]
where $\omega_1=\alpha_1\beta_1/m_e$ and $\omega_2=\alpha_2\beta_2/m_e$ with
\[
\alpha_1^2=\frac{\Omega_1^2+3\Omega_2^2+\Omega_3^2}{2(\Omega_1^2+\Omega_2^2)}, \beta_1^2 = \frac{1}{4}(3\Omega_1^2+\Omega_2^2+\Omega_3^2),
\]
\[
\alpha_2^2=\frac{3\Omega_1^2+\Omega_2^2-\Omega_3^2}{2(\Omega_1^2+\Omega_2^2)}, \beta_2^2 = \frac{1}{4}(\Omega_1^2+3\Omega_2^2-\Omega_3^2),
\]
\[
\Omega_{1,2}^2=m_e^2(\omega_{x,y}^2+\frac{1}{4}\omega_c^2)
\]
\[
\Omega_3^2=[(\Omega_1^2-\Omega_2^2)^2+2m_e^2\omega_c^2(\Omega_1^2+\Omega_2^2)]^{1/2}
\]
\[
\omega_c=eB/m_ec
\]
Using the same analysis as before and assuming that the ellipsoidal dot elongates with increasing hole number, resulting in $\omega_{1,0}>\omega_{1,1}$ and $\omega_{2,0}>\omega_{2,1}$, the calculation (not shown here due to its length) reveals a $B^2$ dependence additionally to $\mp\frac{1}{2}\mu_B B (g^*_1-g^*_0)$.

To account for a linear variation in the effective $g$-factor, which is visible in the residuals of $\uparrow_\text{o1}-\downarrow_\text{o2}$, it is necessary to go beyond the simple Zeeman Hamiltonian and consider the full Luttinger--Kohn (LK) Hamiltonian. In particular, the off-diagonal elements of the Zeeman Hamiltonian for a perpendicular magnetic field $\vec{B} = B\hat{z}$ are given by:

\[
H_Z = 2\mu_B\left(\kappa B J_z + q B - J_z^3\right),
\]

where $\kappa$ and $q$ are the isotropic and cubic Zeeman parameters, respectively. For states with angular momentum difference $\Delta l = 1$, such as the heavy-hole (HH) state $\ket{HH+} = \ket{3/2, +3/2}$ and the light-hole (LH) state $\ket{LH+} = \ket{3/2, +1/2}$, the spherical approximation does not allow coupling via orbital angular momentum. However, the cubic Zeeman term introduces non-zero matrix elements such as:

\[
\bra{3/2, +1/2} J_z^3 \ket{3/2, +3/2} \neq 0,
\]

indicating that heavy-hole-light-hole mixing can occur through this term.

To capture this mixing more accurately, we consider the full Luttinger--Kohn Hamiltonian:

\[
H = \frac{\hbar^2}{2m_0} \left[ \left(\gamma_1 + \frac{5}{2}\gamma_2\right)k^2 - 2\gamma_2(\vec{k} \cdot \vec{J})^2 \right],
\]

where $\gamma_1$ and $\gamma_2$ are Luttinger parameters, and $\vec{J}$ is the total angular momentum operator for $j = 3/2$. Applying the Peierls substitution $\vec{k} \rightarrow \vec{k} + \frac{e}{\hbar} \vec{A}$ with the symmetric gauge $\vec{A} = \frac{B}{2}(-y, x, 0)$, we obtain:

\[
k_x \rightarrow k_x - \frac{eB_z}{2\hbar}y, \quad k_y \rightarrow k_y + \frac{eB_z}{2\hbar}x.
\]

The heavy-hole-light-hole mixing (HH-LH) arises from the off-diagonal terms in the LK Hamiltonian, which are proportional to:

\[
H_{HH-LH} \propto k_+^2 J_-^2 + k_-^2 J_+^2,
\]

with $k_\pm = k_x \pm i k_y$ and $J_\pm = J_x \pm i J_y$. Substituting the magnetic field dependence into $k_+$, we find:

\[
k_+ = k_+^{(0)} + \frac{eB_z}{2\hbar}(-y + ix),
\]

so that $k_+^2$ contains terms linear in $B$. Therefore, the heavy-hole-light-hole mixing leads to a linear modification of the effective $g$-factor.

\section{\label{sec:STcrossing} Discontinuity}
\begin{figure}
    \includegraphics[width=\linewidth]{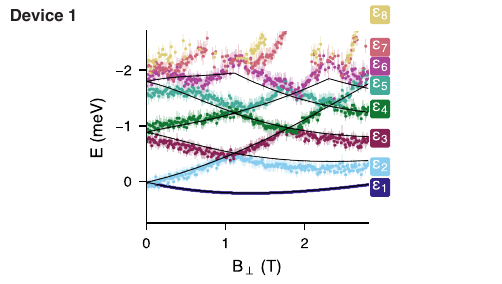}
    \caption{\textbf{Fock-Darwin simulation of data from \figref{fig:1}e} using $m^*=\num{0.09}m_e$, $\hbar\omega_x=\SI{0.9}{\milli e\volt}$, $\hbar\omega_y=\SI{2}{\milli e\volt}$ and $g=\num{11.12}$}
    \label{fig:supplement_fockdarwinData}
\end{figure}

\begin{figure}
    \includegraphics[width=\linewidth]{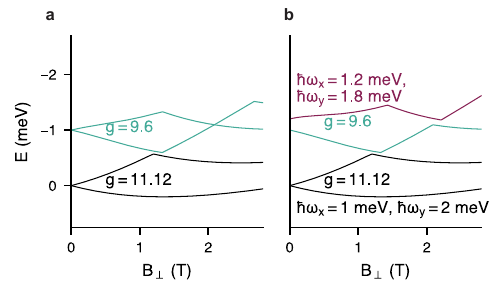}
    \caption{\textbf{Fock-Darwin simulation of quantum dot with decreased $g$-factor and increased anisotropy with hole number.} \textbf{(a)} Simulated Zeeman-split energy levels with varying $g$-factors, indicated by color. The confinement potential is given by $\hbar\omega_x=\SI{1}{meV}$, $\hbar\omega_y=\SI{2}{meV}$ and the effective mass is $m^*=0.09$ \textbf{(b)} The same $g$-factors as in panel a are used, but the confinement potential is modified for the fourth orbital level to reflect variation of the anisotropy. The first three levels retain the original confinement parameters.}
    \label{fig:supplement_discontinuity_Fock}
\end{figure}

\begin{figure}
    \includegraphics[width=\linewidth]{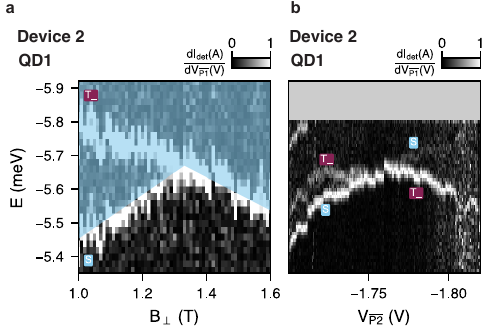}
    \caption{\textbf{Zoom-in of \RefFigtwo d at the singlet–triplet crossing.} 
\textbf{(a)}, Measured as a function of magnetic field. 
\textbf{(b)}, Measured via voltage-tunable $g$-factor by varying the plunger gate of the second, which is the unoccupied QD2 at $N = 0$.}
    \label{fig:supplement_discontinuity}
\end{figure}

To investigate the influence of orbital effects, hole-hole interactions, and confinement on the energy spectrum, we compare the addition spectrum (CBAS) with the single-particle excitation spectrum (PESS), both extracted from the same magnetospectroscopy dataset. The CBAS, shown in \figref{fig:fig2}a, is derived from the ground states for different hole numbers in \figref{fig:fig2}d using the previously described method. For transitions $N = 1 \leftrightarrow 2$ and $N = 2 \leftrightarrow 3$, the CBAS ground-state energies exhibit a discontinuity near $B_\perp = \SI{1.49}{T}$, indicated by a yellow dashed line. 
This feature is also visible when overlaying the data from \figref{fig:1}e with an anisotropic Fock-Darwin model as described in Ref. ~\cite{Madhav94}, shown in \figref{fig:supplement_fockdarwinData}. While the model captures the general trend of the energy levels, deviations occur at level crossings and for higher excited states. 

The observed discontinuity likely originates from changes in the confinement potential as hole occupancy increases, modifying orbital energies and shifting magnetic-field crossings. 
To gain insight into the origin of the observed discontinuities, we simulate the system using an anisotropic Fock–Darwin model, following the approach of Madhav and Chakraborty~\cite{Madhav94}. By varying the $g$-factors for the different Zeeman-split pairs (see \figref{fig:supplement_discontinuity_Fock}a), as measured for device 1 (see \figref{fig:1}{e}), a discontinuity emerges at the first level crossing. This arises because the two crossing states have different $g$-factors and therefore also different $\epsilon_1$, while the offset is always subtracted starting at \SI{0}{meV}. This requires adapting the chemical potential description from Ref.~\cite{Kouwenhoven_2001} to account for level-specific energies for each hole $N$:
\begin{align*}
&\mu(1) = E_{0,0}(N=0)=\epsilon_1 \\
&\mu(2) = E_c + E_{0,0}(N=1)
\end{align*}
In contrast, the next higher crossing does not exhibit a discontinuity, as both involved states share the same $g$-factor.
However, as shown in Fig. \figref{fig:1}{e} and \figref{fig:supplement_fockdarwinData}, discontinuities are also present at higher crossings (for each level, the first level is adapted to the parameters). These can be explained by changes in the anisotropy of the confinement potential, as illustrated in \figref{fig:supplement_discontinuity_Fock}b. Due to the number of free parameters for each energy level ($g(N), \omega_x(N), \omega_y(N), m^*(N)$), the risk of overfitting is high and offers limited physical insights. Additionally, spin–orbit interactions may contribute to such discontinuities, as discussed by Bulaev and Loss~\cite{Bulaev05}. These findings highlight the sensitivity of the energy spectrum to both orbital configuration and spin-dependent interactions, emphasizing the challenge for simple models. Therefore this shows the need for detailed modeling to interpret experimental observations.

To investigate the discontinuity observed in the addition spectrum further, we examine the PESS data near the singlet–triplet crossover (\figref{fig:supplement_discontinuity}a). Here, the discontinuity is not observed in the PESS data but appears clearly in the CBAS measurements, where each energy level corresponds to the addition of a hole. This contrast highlights the role of many-body interactions and confinement changes upon hole addition, which are more pronounced in the addition spectrum. Furthermore, no anticrossing is visible in the PESS data. To extract an upper bound on the Rashba spin–orbit coupling gap at the degeneracy point~\cite{Bulaev05,Bulaev05_2}, we determine the full width at half maximum (FWHM) of the transition line to be $\SI{52.7 \pm 18.1}{\micro eV}$.

In a slightly modified configuration of tunnel rates, the singlet branch beyond the crossing becomes faintly visible in \figref{fig:supplement_discontinuity}b. Here, the crossing is measured by tuning both plunger gates, demonstrating that the $g$-factor voltage tunability is sufficient to traverse the crossover. This suggests that hole–hole interactions and modifications to the confinement potential upon hole addition significantly influence the energy spectrum—effects that are less pronounced in the PESS data.

\section{\label{sec:Fig2supplement_tunnelrate} Tunnel rate measurement for PESS}
\begin{figure}
    \includegraphics[width=\linewidth]{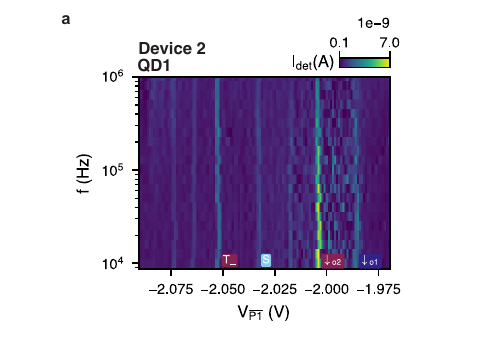}
    \caption{\textbf{Tunnel rate extraction from AC pulse spectroscopy with $B_\perp=$\SI{2.5}{\tesla}, $\delta \vPone =$\SI{0.2}{\volt}, $\vPtwo =$ \SI{-1.6}{\volt}.} 
    AC pulse frequency is swept against the DC gate voltage applied to QD1 of device 2 to probe the tunnel rate.}
    \label{fig:supplement_fig2}
\end{figure}

To measure the tunnel rates of different ground and excited states, we sweep the AC pulse frequency while monitoring the response as a function of DC gate voltage (\figref{fig:supplement_fig2}). Ground and excited states appear as lines in the data, which indicate tunneling events between the quantum dot and the reservoir. If the tunnel rates were lower than the pulse frequency (repetition rate), these lines would disappear. Since the lines remain visible down to the lowest measurable frequency (\SI{1}{\kilo\hertz}), we conclude that the tunnel rate is above the detection limit but still sufficient to observe transitions.

\section{\label{sec:sup_EnergyComparison} Comparison of energy scales}
To understand the impact of orbital in comparison to the spin, we calculate the energy scales within the constant interaction model.
Zeeman energy, $E_Z$, is given by:
\[
E_Z=g^*\mu_B B = g^* \frac{e\hbar B}{2 m_e}
\]
where $m_e$ is the electron mass.
The orbital energy, $E_\text{orb}$ scale is given by:
\[
E_\text{orb}=\hbar \omega_c =\frac{\hbar e B}{m^*}
\]
where $m^*=0.09\pm0.002m_e$~\cite{sammak19} is the effective mass.
Using $g^*=10$, the proportion is 
\[
\frac{E_Z}{E_\text{orb}}=0.45
\]
Therefore, the energy scales are comparable and it is a strong spin-orbit coupled system.

 \section{\label{sec:CBAS_fulldata} Addition spectra in dependence of $V_{\overline{B12}}$}
 \begin{figure*}
    \includegraphics[width=\textwidth]{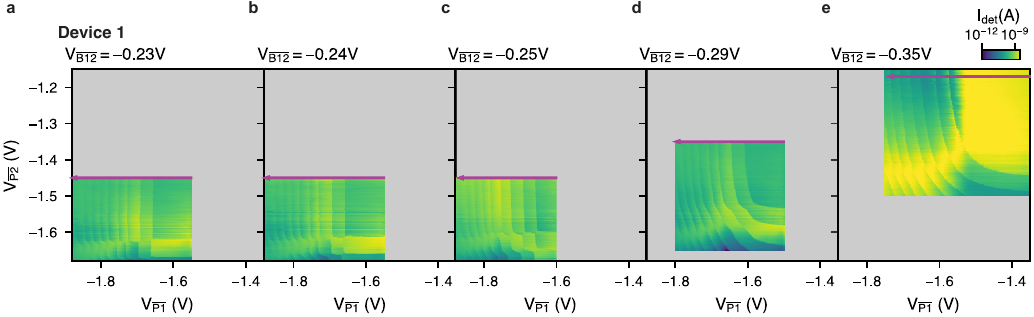}
    \caption{\textbf{Charge stability diagrams for different barrier voltages $V_{\overline{B{12}}}$ on device 1.} 
The purple arrows indicate the voltage configurations where the data for the addition spectrum was acquired.
    \textbf{(a)} $V_{\overline{B12}} =$\SI{-0.23}{\volt}
    \textbf{(b)} $V_{\overline{B12}} =$\SI{-0.24}{\volt}
    \textbf{(c)} $V_{\overline{B12}} =$\SI{-0.25}{\volt}
    \textbf{(d)} $V_{\overline{B12}} =$\SI{-0.29}{\volt}
    \textbf{(e)} $V_{\overline{B12}} =$\SI{-0.35}{\volt}}
    \label{fig:CBAS_full_CSD}
\end{figure*}
In \figref{fig:CBAS_full_CSD}, the charge stability diagrams for the double quantum dot system are shown for decreasing values of $V_{\overline{B{12}}}$. As $V_{\overline{B{12}}}$ becomes more negative, the tunnel coupling between the two quantum dots increases, as evidenced by the broadening at the triple points. 
\begin{figure}
    \centering
    \includegraphics[width=\linewidth]{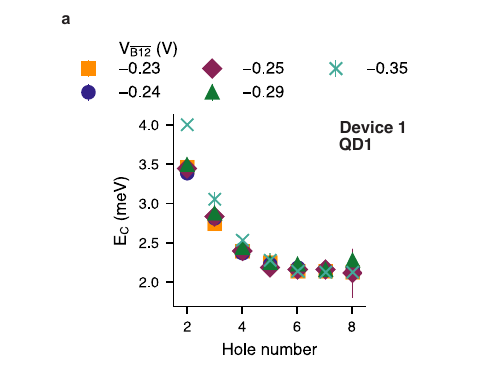}
    \caption{\textbf{(a)} Extracted charging energies (gray) for different barrier voltages $V_{\overline{B{12}}}$ as a function of total dot occupation. Error bars are
smaller than the marker size.}
    \label{fig:EC_VB12}
\end{figure}

\begin{figure*}
    \includegraphics[width=\textwidth]{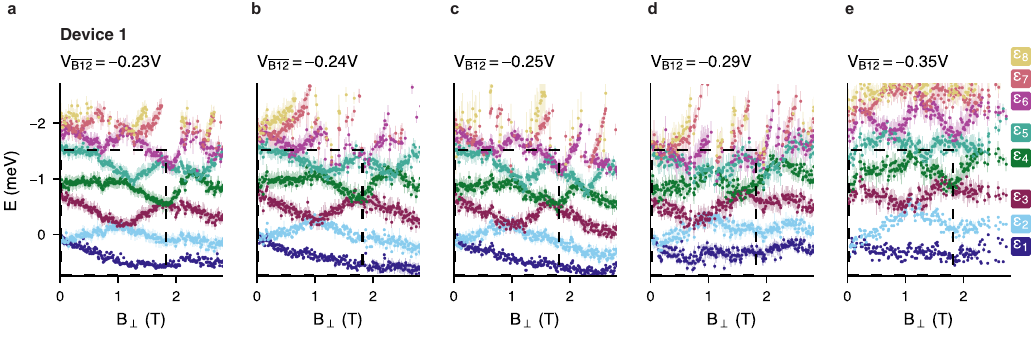}
    \caption{\textbf{Addition spectra for different barrier voltages $V_{\overline{B{12}}}$.} 
The data were obtained by subtracting the charging energy $E_C$ (\figref{fig:EC_VB12}a) from the addition energies on device~1. 
The dashed square highlights the region used for comparison with \figref{fig:1}e, \figref{fig:fig2}a and between the panels.
    \textbf{(a)} $V_{\overline{B12}} =$\SI{-0.23}{\volt}
    \textbf{(b)} $V_{\overline{B12}} =$\SI{-0.24}{\volt}
    \textbf{(c)} $V_{\overline{B12}} =$\SI{-0.25}{\volt}
    \textbf{(d)} $V_{\overline{B12}} =$\SI{-0.29}{\volt}
    \textbf{(e)} $V_{\overline{B12}} =$\SI{-0.35}{\volt}}
    \label{fig:CBAS_full}
\end{figure*}

To extract the ellipticity of QD1, we analyze the energy difference $\epsilon_5 - \epsilon_4$ at $B = \SI{0}{\tesla}$ for decreasing values of $V_{\overline{B{12}}}$. The results are plotted in \RefFigthree, where, for $N = 4$, the single-particle energy is smaller at $V_{\overline{B{12}}} = \SI{-0.35}{\volt}$ compared to higher voltages. This indicates that the quantum dot becomes less elliptical as $V_{\overline{B{12}}}$ is made more negative, with the confinement potential approaching a less ellipsoidal confinement potential shape at $V_{\overline{B{12}}} = \SI{-0.35}{\volt}$ relative to $V_{\overline{B{12}}} = \SI{-0.23}{\volt}$. A discontinuity is visible at level crossings in all spectra. The voltage tunability of the $g$-factor is directly evident from the crossing of $\epsilon_2$ and $\epsilon_3$, which occurs at different magnetic fields for different values of $V_{\overline{B{12}}}$. This observation indicates that, although the overall confinement potential shape of the quantum dot does not change drastically, the $g$-factor is significantly affected by the gate voltage.

\section{\label{subsec:dot2} Pulsed excited-state spectroscopy of dot under $V_{\overline{P2}}$}

\begin{figure}[htbp]
    \includegraphics[width=\linewidth]{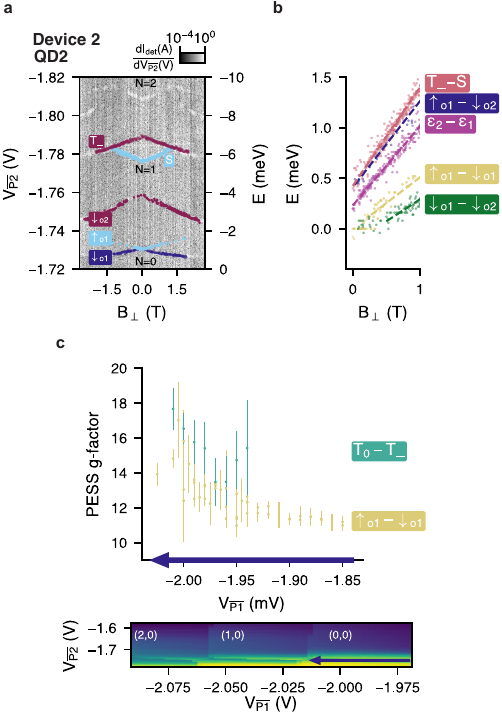}
    \caption{\textbf{Pulsed excited-state  spectroscopy and $g$-factor extraction for QD2.} 
    \textbf{(a)} PESS measurement of QD2. $\diff I_\textnormal{det}/\diff V_{\overline{\textnormal{P1}}}$ as a function of $\vPtwo$ and $B_\perp$ at fixed $\delta \vPtwo$. The hole number $N$ is labeled, and regions sensitive to excited-state spectra are highlighted. Extracted energy levels are overlaid as colored lines.
    \textbf{(b)} Extracted spin-split energy pairs from PESS are analyzed to determine the CBAS $g$-factor for hole occupations $N = 1$ and $N = 2$, following the same procedure as in the CBAS method. Additionally, PESS $g$-factors are extracted from the spin-split excited-state transitions including orbital contribution ($\downarrow_\text{o1}$–$\downarrow_\text{o2}$, $\uparrow_\text{o1}$–$\downarrow_\text{o1}$, $T_-$–$S$).
    \textbf{(c)} PESS $g$-factor of QD2 as a function of the plunger gate voltage applied to the neighboring quantum dot. The lower plot shows the corresponding charge stability diagram, with arrows indicating the direction of plunger gate tuning. The charge configuration is labeled as $(N,M)$. 
    }  
    
    \label{fig:fig8}
\end{figure}

PESS measurements were also performed on the second quantum dot (QD2) in the double quantum dot (DQD) system of device 2. However, due to the complex voltage landscape near QD2—likely influenced by the proximity of the charge sensor, we were unable to optimize the tunnel couplings to resolve additional excited states beyond those shown in \figref{fig:fig8}a. From the visible spin-split pair, we extract the differences of the levels at $B_\perp=\SI{0}{\tesla}$ $\Delta \text{orb}_{N_{0\leftrightarrow1}} = \SI{2.76}{meV}$, $\Delta \text{orb}_{N_{1\leftrightarrow2}} = \SI{1.36}{meV}$. In comparison to QD1, the differences are larger by \num{7.81}\% for $\Delta \text{orb}_{N_{0\leftrightarrow1}}$ and \num{27.1}\% for $\Delta \text{orb}_{N_{1\leftrightarrow2}}$. This indicates within the Fock-Darwin model, that the confinement is tighter for QD2.

Furthermore, we investigate the $g$-factors of QD2, which are extracted from \figref{fig:fig8}b and listed in \cref{tab:g_supp}. 
Compared to QD1, the PESS $g^{{\uparrow_\text{o1}}-{\downarrow_\text{o1}}}$-factor is reduced, while the PESS $g^{{\uparrow_\text{o1}}-{\downarrow_\text{o2}}}$-factor and PESS $g^{{S}-{T_0}}$-factor are significantly enhanced. Additionally, using the CBAS method to extract the orbital energy difference yields $g^{\epsilon_2 - \epsilon_1} = \num{13.53 \pm 0.18}$, which aligns more closely with measurements taken at more negative barrier voltages as seen in \figref{fig:fig3}b. These observations suggest that, despite nominally identical gate geometries, the two dots exhibit distinct confinement potentials, likely due to local strain variations or differing proximities to the charge sensor, which result in a modified potential landscape.

Additionally, we shifted the wavefunction toward the interdot charge transition using the plunger gate $\vPone$, as shown in the bottom panel of Fig.~\ref{fig:fig8}c, to investigate the tunability of the pure-spin PESS $g$-factor. Near the transition, both $g^{\uparrow_\text{o1}-\downarrow_\text{o1}}$ and $g^{T_0-T_{-}}$ exhibit an apparent increase. A similar trend for $g^{T_0-T_{-}}$ in QD1 is observed in Fig.~\ref{fig:fig3}c, indicating that the $g$-factor changes continuously as the wavefunction shifts from one plunger gate to the other.

\begin {table}[h!]
\caption {Extracted absolute $g$-factors with $3\sigma$-error bars from CBAS and PESS for dot under $V_{\overline{P2}}$ of device 2, where N is the hole number. Additionally, the orbital number investigated in the analysis are indicated.} \label{tab:g_supp} 
\begin{center}
\begin{tabular}{|c|c|c|c|c|}
\hline
\multicolumn{5}{|c|}{\textbf{Device 2 - QD2}} \\
\hline
& method &  & $N$ & orbital \\
\hline

$g^{\epsilon_2 - \epsilon_1}$ & CBAS &\num{13.53 \pm 0.54}&\num{1},\num{2} & \num{1} \\
\hline
\multicolumn{5}{|c|}{bare spin} \\
\hline
$g^{\uparrow_\text{o1} - \downarrow_\text{o1}}$ &PESS &\num{8.8\pm0.51}&\num{1} & \num{1}\\
\hline
\multicolumn{5}{|c|}{spin \& orbital} \\
\hline
$g^{\uparrow_\text{o1} - \downarrow_\text{o2}}$ &PESS &\num{14.72\pm0.66}&\num{1} & \num{1},\num{2}\\
$g^{S - T_{-} }$ &PESS&\num{16.85\pm0.63}&  \num{2}& \num{1},\num{2}\\

\hline
\end{tabular}
\end{center}
\end {table}

\section{\label{sec:tiltedfield} Tilted magnetic fields}
Complementary to the magnetic field strength dependence, we also investigated the anisotropy of the $g$-factor, as reported in previous studies~\cite{Hendrickxsweetspot24,brickson2024usinghighfidelitynumericalmodel}. For this, magnetospectroscopy measurements were performed as a function of the magnetic field angle $\theta$, while maintaining a constant field magnitude of $|B| = \SI{1}{\tesla}$. Two in-plane field orientations were considered: $\phi = 0^\circ$ and $\phi = 90^\circ$, as indicated by the lines in \RefFigthree a and b. The analysis followed the procedure described in the main text. We assumed zero Zeeman splitting at $|B| = \SI{0}{\tesla}$, consistent with physical expectations. The $g$-factor was thus determined from the slope of the energy splitting. 

\begin{figure*} 
    \includegraphics[width=\textwidth]{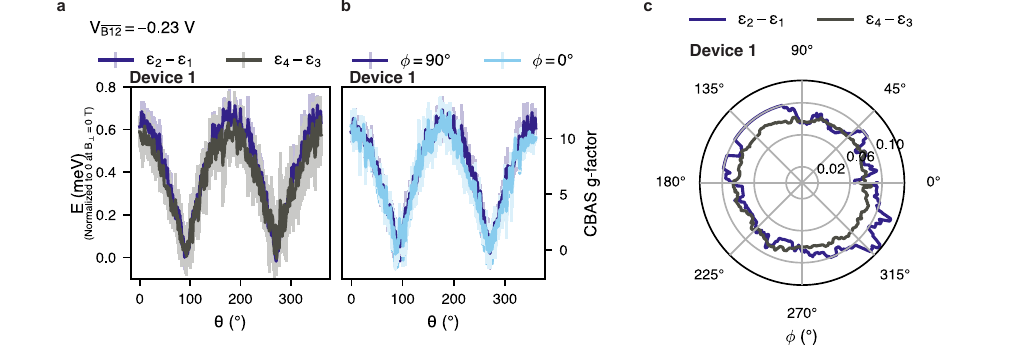}
    \caption{\textbf{Tilted magnetic field magnetospectroscopy and CBAS $g$-factor anisotropy.} 
    Measurements are performed at $V_{\overline{\mathrm{B{12}}}} = \SI{-0.23}{V}$ and $|B| = \SI{1}{\tesla}$. The CBAS $g$-factor is extracted by subtracting spin-split energy levels and fitting the slope from the origin ($B = \SI{0}{\tesla}$, spin splitting = 0) for two spin pairs: $(\epsilon_2 - \epsilon_1)$ and $(\epsilon_4 - \epsilon_3)$.
    \textbf{(a)} CBAS $g$-factor anisotropy as a function of hole number at fixed azimuthal angle $\phi = 0\si{\degree}$. 
    \textbf{(b)} CBAS $g$-factor anisotropy for two different values of $\phi$ for the $N = 1 \leftrightarrow 0$ transition. 
    \textbf{(c)} In-plane CBAS $g$-factor for different hole numbers measured at polar angle $\theta = 90\si{\degree}$.}
    {Panels a and b share the same y-axes for direct comparison.}
    \label{fig:fig_tilted}
\end{figure*}
The results, measured with CBAS for both in-plane angles overlapped within the experimental uncertainty of approximately \SI{2}{\degree}.
The $g$-factor reaches its maximum when the magnetic field is oriented perpendicular to the quantum well plane ($\theta = 0^\circ$ or $180^\circ$), , indicating that any tilt of the $g$-tensor is below our detection threshold (see \figref{fig:fig_tilted}b). 
In contrast, the in-plane $g$-factor is significantly smaller, consistent with previous reports~\cite{Hendrickxsweetspot24}, and reaches values around \num{0.06}, which is in the same range as previously reported values in literature ~\cite{Hendrickx20,Wang_Hopping_spins,Hendrickxsweetspot24}.
These measurements were also performed at a fixed magnetic field magnitude of $|B| = \SI{1}{\tesla}$. A slight anisotropy is observed in the in-plane $g$-factor for the orbital splitting $g^{\epsilon_2 - \epsilon_1}$. However, due to the relatively large experimental uncertainty and the possibility of misalignment between the device and the magnet’s coordinate system, this anisotropy cannot be conclusively resolved. As such, more precise techniques, such as qubit spectroscopy, would be required to detect the in-plane anisotropy reported in earlier studies~\cite{Hendrickxsweetspot24}.

\begin{figure*}
    \includegraphics[width=\textwidth]{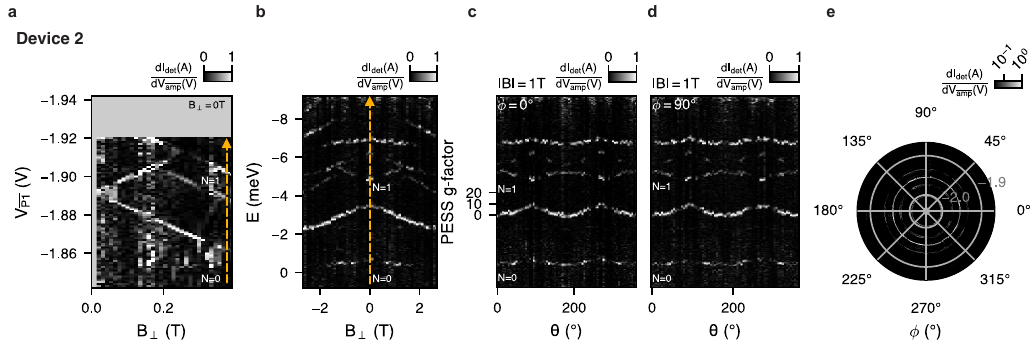}
    \caption{\textbf{PESS in dependence of tilted magnetic field. } 
\textbf{(a)} Derivative of the charge sensor current, $|dI_{\mathrm{SD}}/dV_{\overline{\mathrm{P1}}}|$, plotted as a function of pulse amplitude $\delta V_{\overline{\mathrm{P1}}}$ and dc gate voltage $\vPone$ at $B_\perp = 0$. 
\textbf{(b)} $|dI_{\mathrm{SD}}/dV_{\overline{\mathrm{P1}}}|$ as a function of $\vPone$ and $B_\perp$ at fixed $\delta \vPone$. Hole number $N$ is indicated. 
\textbf{(c)} PESS $g$-factor anisotropy measured at $|B| = \SI{1}{\tesla}$ for azimuthal angle $\phi = 0\si{\degree}$.  The $g$-factor is extracted from the slope of the spin-split energy versus magnetic field, extrapolated to the origin ($B = \SI{0}{\tesla}$, spin splitting = 0). The $y$-axis is consistent with panels \textbf{a} and \textbf{b}.  
\textbf{(d)} Same as panel \textbf{c}, but for $\phi = 90\si{\degree}$. 
\textbf{(e)} In-plane anisotropy of the PESS $g$-factor measured at polar angle $\theta = 0^\circ$. Labels indicate the corresponding $\vPone$ values.}
{Panels (a–d) share the same y-axes for direct comparison.}
    \label{fig:fig9_tilted}
\end{figure*}

In addition, we investigated the single-particle excitation spectrum in a tilted magnetic field using the PESS method, as shown in \figref{fig:fig9_tilted}a,b. The first excited state exhibits a strongly anisotropic $g$-factor, varying between approximately from \num{0} to \num{10} depending on the magnetic field orientation (\figref{fig:fig9_tilted}c,d). However, no clear anisotropy is observed in the in-plane $g$-factor using this method (\figref{fig:fig9_tilted}e), which may be attributed to a slight tilt of the device relative to the magnet’s coordinate system.
These findings are consistent with those obtained via the CBAS method, confirming that both approaches yield qualitatively similar conclusions. The quantitative differences arise from the nature of the two techniques: PESS probes single-hole excited states, whereas CBAS involves many-body effects, as discussed in the main text.

\end{document}